\documentclass[12pt]{article}

\usepackage[utf8]{inputenc}
\usepackage{color}
\usepackage{amsfonts,amsmath}
\usepackage[mathscr]{eucal}
\usepackage{amssymb}
\usepackage{cite}
\usepackage{bm}
\usepackage{bbold}
\usepackage{amsthm}
\usepackage{graphicx,subfigure}
\usepackage{epsf}

\newcommand{\be}{\begin{eqnarray}}
\newcommand{\ee}{\end{eqnarray}}
\newcommand{\nn}{\nonumber \\}
\newcommand{\lb}{\label}
\newcommand{\p}[1]{(\ref{#1})}
\newcommand{\ga}{\lower.7ex\hbox{$
\;\stackrel{\textstyle>}{\sim}\;$}}

\newcommand{\la}{\lower.7ex\hbox{$
\;\stackrel{\textstyle<}{\sim}\;$}}

\begin{document}

\begin{center}
{\LARGE\bf Modified Korteweg-de Vries Equation as a System with Benign Ghosts\footnote{A talk at the AAMP conference, Prague, September 2021, to be published in {\it Acta Polytechnica}.}}

\vspace{15mm}

 {\large\bf Andrei~Smilga} 
 \vspace{0.5cm}

{\it SUBATECH, Universit\'e de Nantes,}\\
{\it 4 rue Alfred Kastler, BP 20722, Nantes 44307, France;}\\
\vspace{0.1cm}

{\tt smilga@subatech.in2p3.fr}\\

\end{center}

\vspace{1cm}

\begin{abstract}
We consider the modified Korteweg-de Vries equation,
$ u_{xxx} + 6u^2 u_x + u_t \ =\ 0 $,
and explore its dynamics in {\it spatial} direction. Higher $x$ derivatives 
bring about the {\it ghosts}. We argue that these ghosts are benign, i.e. the classical dynamics of this system does not involve a blow-up. This probably means that also the associated quantum problem is well defined.

\end{abstract}

\newpage

\section{Introduction}

A system with ghosts is by definition a system where the quantum Hamiltonian has no ground state so that its spectrum involves the states with arbitrarily low and arbitrarily high energies. In particular, all non-degenerate theories with higher derivatives in the Lagrangian (but not only them!) involve ghosts. The ghosts show up there already at the classical level: the Ostrogradsky Hamiltonians of higher derivative systems \cite{Ostro} include the linear in momenta terms  and are thus not positive definite \cite{Woodard}. This brings about the ghosts in the quantum problem \cite{est,obzor}.

\begin{figure} [ht!]
      \begin{center}
    \includegraphics[width=.5\textwidth]{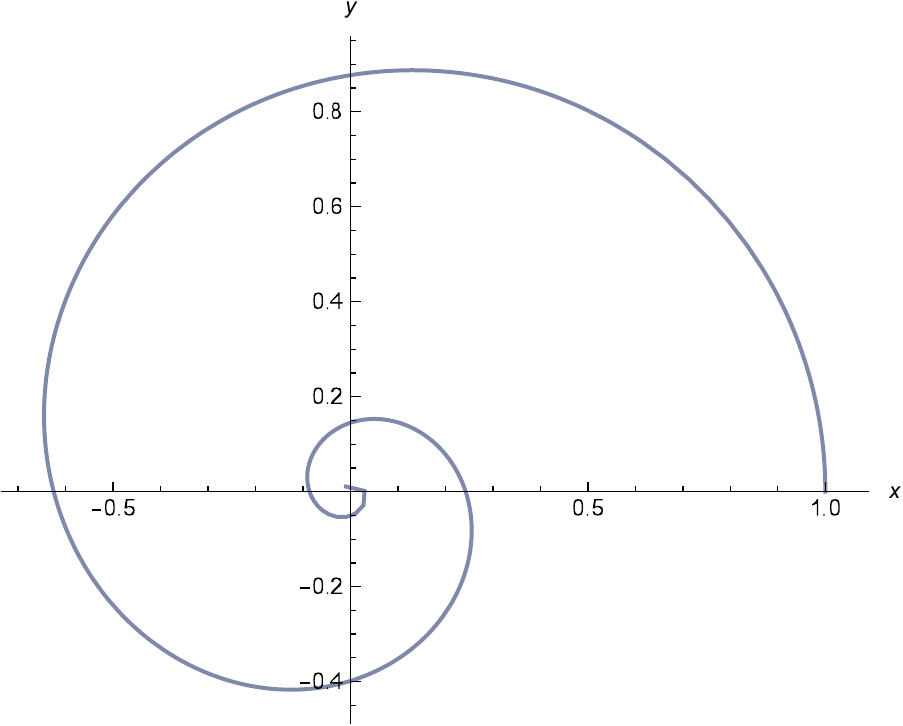}                  
     \end{center}
    \caption{Falling on the center for the Hamiltonian \p{H-center} with $m=1$ and $\kappa = .05$. The energy is slightly negative. The particles with positive energies escape to infinity.}        
 \label{spiral}
    \end{figure}

In many cases, ghost-ridden systems are sick --- the Schr\"odinger problem is not well posed and unitarity is violated. Probably, the simplest example of such a system is a system with the Hamiltonian
 describing the 3-dimensional motion of a particle in an attractive $\frac1{r^2}$ potential:
    \be
       \lb{H-center}
       H \ =\ \frac {\vec{p}^2}{2m} - \frac {\kappa} {r^2} \, .
        \ee
 Classically, for certain initial conditions, the particle falls to the center
 in a finite time, as is shown in Fig. \ref{spiral}. 

 The quantum dynamics of this system depends  on the value of ${\kappa}$. If $m{\kappa} < 1/8$, the ground state
 exists and unitarity is preserved.  If $m{\kappa} > 1/8$, the spectrum is not bounded from below and, what is worse, the quantum problem cannot
 be well posed until the singularity at the origin is smoothed out \cite{Popov}. 
 One can say that for $m{\kappa} < 1/8$ quantum fluctuations cope successfully with the attractive force of the potential and prevent the system from collapsing.
 
The latter example suggests that quantum fluctuations can only make a 
ghost-ridden system better, not worse. 
We therefore {\it conjecture} that, if the classical dynamics of the system is benign, i.e. the system does not run into a singularity in finite time,\footnote{We still call a system benign if it runs into a singularity at $t = \infty$. Such systems have well-defined quantum dynamics. This refers e.g. to the problem of motion in a uniform electric field (see e.g. \cite{Landau}, \S24) and also to the inversed oscillator with the Hamiltonian $H = (p^2 - x^2)/2$. In the latter problem, the classical trajectories $x(t)$ grow exponentially with time, but the quantum problem is still benign (see e.g. \cite{Robert1}, Ch. 3, corollary 13). The spectrum in this case is continuous, as it is for the uniform field problem.}
its quantum dynamics will also be  benign, irrespectively of whether the spectrum has, or does not have, a bottom.

This all refers to ordinary mechanical or field theory systems, where energy is conserved and the notion of  Hamiltonian exists. The ghosts in gravity (especially, in higher-derivative gravity) is a special issue that we are not discussing here.

Besides malignant ghost-ridden systems, of which the system \p{H-center} with $m{\kappa} > 1/8$ represents an example,  there also are many systems with ghosts which are {\it benign} --- unitarity is preserved and the quantum Hamiltonian is self-adjoint with a well-defined real spectrum.
To begin with, such is the famous {\it Pais-Uhlenbeck oscillator} \cite{PU} --- a higher derivative system with the Lagrangian 
\be
      \lb{LPU}
      L \ =\ \frac 12 \left[ \ddot{x}^2 - (\omega_1^2 + \omega_2^2) \dot{x}^2 + \omega_1^2 \omega_2^2 x^2 \right]. 
        \ee  
This system is free, its canonical Hamiltonian can be reduced to the difference of the two oscillator Hamiltonians by a canonical transformation \cite{Mann-Dav}.  The first example of a nontrivial benign ghost system involving nonlinear interactions was built up in \cite{Robert}.
For other such examples, see Refs. \cite{Pavsic,Kovner,duhi-v-pole,Toda-ja,Deffayet}.

In recent \cite{Damour}, we outlined two wide classes of benign ghost systems: {\it (i)} the systems obtained by a variation of ordinary systems and involving, compared to them, a double set of dynamic variables and {\it (ii)} the systems describing geodesic motion over Lorenzian manifolds. In addition, we noticed that the evolution of the modified Korteweg-de Vries (MKdV) system 
\p{MKdV} in the {\it spatial} direction also exhibits a benign ghost dynamics. This report is mostly based on section 4 of Ref. \cite{Damour} that deals with the MKdV dynamics.

     \section{Spatial dynamics of  KdV and MKdV equations}

Consider first the ordinary KdV equation, 
 \be  \lb{KdV}
    u_{xxx} + 6 u u_x  + u_t   = 0\,,
\ee
where $u_x = \partial u/\partial x, u_t = \partial u/\partial t$ etc.)
     It has an infinite number of integrals of motion and is exactly soluble.\footnote{Exact solvability always makes the behaviour of a system more handy. In particular, many mechanical models including benign ghosts, which were mentioned above, are exactly solvable.} The KdV equation is derived  from the field Lagrangian
\be \label{LKdV}
L[\psi(t,x)] \ =\ \frac 12 \psi_{xx}^2 - \psi_x^3 - \frac 12 \psi_t \psi_x\,
\ee
if one denotes $u(t,x) \equiv \psi_x$ after having varied {over $\psi(t,x)$}.
This Lagrangian involves higher spatial derivatives, but not higher time derivatives and does not involve ghosts in the ordinary sense. We can, however, simply 
 {\it rename}  
\be 
\lb{rename}
t \to X,  \qquad x \to T\,,
\ee 
in which case the equation acquires the form 
\be \lb{KdV-rotate}
      u_{TTT} + 6 u u_T  + u_X   = 0
\ee 
and higher time derivatives appear. According to our conjecture, to study the question of whether the quantum Hamiltonian corresponding to the thus rotated Lagrangian \p{LKdV} is Hermitian and unitarity is preserved, it is sufficient to study its classical dynamics:  if it does not involve a blow-up and all classical trajectories exist at all times $T$, one can be sure that the quantum system is also benign.

Note that the  question whether or not blowing up trajectories are present is far from being trivial. The ordinary Cauchy problem for the equation \p{LKdV} consists in setting the initial 
 value of $u(t_0,x)$ at a given time moment, say, $t_0=0$. And we are interested now [staying with Eq. \p{LKdV} and not changing the name of the variables according to \p{rename}]
in the Cauchy problem in $x$ direction. The presence of third spatial derivatives in \p{LKdV} makes it necessary to define at the line $x = x_0$  {\it three} different functions:
$u(t, x_0), u_x(t, x_0)$ and $u_{xx}(t, x_0)$.  The presence of  three arbitrary functions makes  the space of solutions to the spatial Cauchy problem  much larger than for the ordinary Cauchy problem. The solutions  to the latter represent a subset of measure zero in the set of the solutions in the former, and the fact that the solutions to the ordinary Cauchy problem are all benign does not mean that it is also the case for the rotated $x$-directed problem.

And, indeed, for the ordinary KdV equation \p{LKdV} the problem is {\it not} benign. It is best seen if we choose a $t$-independent Ansatz $u(t,x) \to u(x)$ and plug it into \p{KdV}.
The equation is reduced to
\be
\lb{KdV-notime}
\partial_x(u_{xx} + 3u^2) = 0 \qquad \Longrightarrow \qquad u_{xx} + 3u^2 = C\,.
\ee
This equation describes the motion in the cubic potential $V(u) = u^3-Cu$. It has  blow-up solutions. If $C=0$, they read
\be
      \lb{runaway}
      u(x) \ =\ -\frac 2{(x-x_0)^2} \,.
\ee
 However, the situation is completely different for the {\it modified} KdV equation,\footnote{In Ref. \cite{Damour}, we wrote this equation as 
$$ u_{xxx} + 12 \kappa u^2 u_x  + u_t   = 0\,$$
and kept $\kappa$ in  all subsequent formulas. But here we have chosen for simplicity to fix $\kappa = 1/2$.}
\be  
\lb{MKdV}
    u_{xxx} + 6 u^2 u_x  + u_t   = 0\,.
\ee
This equation admits an infinite number of  integrals of motion, as the ordinary KdV equation does. The first three local conservation laws are 
 \be
       \lb{protoEnergy}
       \partial_t u = - \partial_x(u_{xx} + 2 u^3)\,,
       \ee
       \be
       \lb{Energy}
       \partial_t u^2 = -2\partial_x\left[\frac 32 u^4 + u u_{xx} - \frac 12 u_x^2 \right]\,,
       \ee
       
       \be
       \lb{superEnergy}
       &&\partial_t\left(\frac 12 u^4 - \frac 12 u_x^2 \right)  =\nn
 &&\!\!\!\!\!\!\!\!\!\!\!\!\!\!\!\!\!\!\!\partial_x\!\!\left[ u_x(2 u^2 u_x + \frac 12 u_{xxx})
       - \frac 12 u_{xx}^2 - 2 u^3 u_{xx} - 2 u^6  \right]
        \ee 
For the time-independent Ansatz, we obtain instead of \p{KdV-notime}
\be
\lb{MKdV-notime}
\partial_x(u_{xx} + 2u^3) = 0 \qquad \Longrightarrow \qquad u_{xx} + 2u^3 = C\,.
\ee
This describes the motion in a {\it quartic} potential  $V(u) = u^4/2- Cu$. This motion is bounded, the solutions being elliptic functions. 

 This observation presents an argument that the rotated Cauchy problem for the equation \p{MKdV} with arbitrary initial conditions on the line $x = const$ might be benign.

Note that this behaviour is specific for the equation \p{MKdV} with the { positive} sign of the middle term (the so-called {\it focusing} case). Plugging the time-independent Ansatz in the {\it defocusing} MKdV equation,\footnote{The coefficient 6 is a convention. It can be changed by rescaling $t$ and $x$. But the sign stays invariant under rescaling.}
\be  \lb{MKdVm}
    u_{xxx} - 6 u^2 u_x  + u_t   = 0\,,
\ee
the problem would be reduced to the motion in the potential  $V(u) = -u^4/2- Cu$ characterized by a blow-up. This conforms to the well-known fact that any solution $u(t,x)$ of the  ordinary KdV equation is related to a solution $v(t,x)$ of the   defocusing  MKdV equation by the Miura transformation, 
\be
         \lb{Miura}
          u \ =\ -(v^2  +  v_x)\,.
           \ee

    A different (though related) analytic argument indicating the absence of real blow-up solutions for the focusing MKdV equation
comes from the analysis of its scaling properties. It is easily seen 
that Eq. \p{MKdV} is invariant under the rescalings $u= \lambda_u \bar u$, $x= \lambda_x \bar x$,
 $t= \lambda_t \bar t$ 
if
\be
\lambda_t = \lambda_x^3, \qquad  \lambda_u = \lambda_x^{-1}\,.
\ee
The quantities $ x u$ and $ x/t^{1/3}$ are invariant under these rescalings. Using also
the space and time translational invariance of the MKdV equation, we can 
look for scaling solutions of the type
\be \lb{scale-Ans}
u(t,x)= \frac{1}{[3(t-t_0)]^{1/3}} w(z)\,,
\ee
where
\be
z \ =\ \frac {x-x_0}{[3(t-t_0)]^{1/3}} \,.
\ee
Inserting the ansatz \p{scale-Ans} in Eq. \p{MKdV}, one easily verifies that the function $w(z)$
 satisfies the equation
\be
0=w'''+ (6w^2 -z) w'- w=\frac{d}{dz}\! \left[  w'' +  2w^3 - z w  \right]
\ee
Denoting as $ C$ the constant value of the bracket in the last right-hand side, we conclude that $w(z)$ satisfies a second-order equation,
\be \label{P2}
 w'' = -  2w^3 + z w  + C \,.
\ee
For the equation \p{MKdVm}, the same analysis would give the equation
\be \label{P2m}
 w'' = 2w^3 + z w  + C \,.
\ee
These are Painlev\'e II equations \cite{Pain}. In general, Painlev\'e equations  have  pole singularities. And indeed,
a local analysis of Eq. \p{P2m} (keeping the leading-order terms
 $ w'' \approx 2w^3$)
shows that \p{P2m} admits simple poles, $w(z) \approx \pm 1/(z-z_0)$. The existence of a
real simple pole at $z=z_0$ would then correspond to a singular (blow-up) behavior of $u(t,x)$ of the form 
$u(t,x) \propto \left[x-x_0 - z_0  [3 (t-t_0)]^{1/3} \right]^{-1}$. But for the equation \p{P2} 
[and hence for \p{MKdV}], the singularities are absent.

\begin{figure} [ht!]
      \begin{center}
    \includegraphics[width=.5\textwidth]{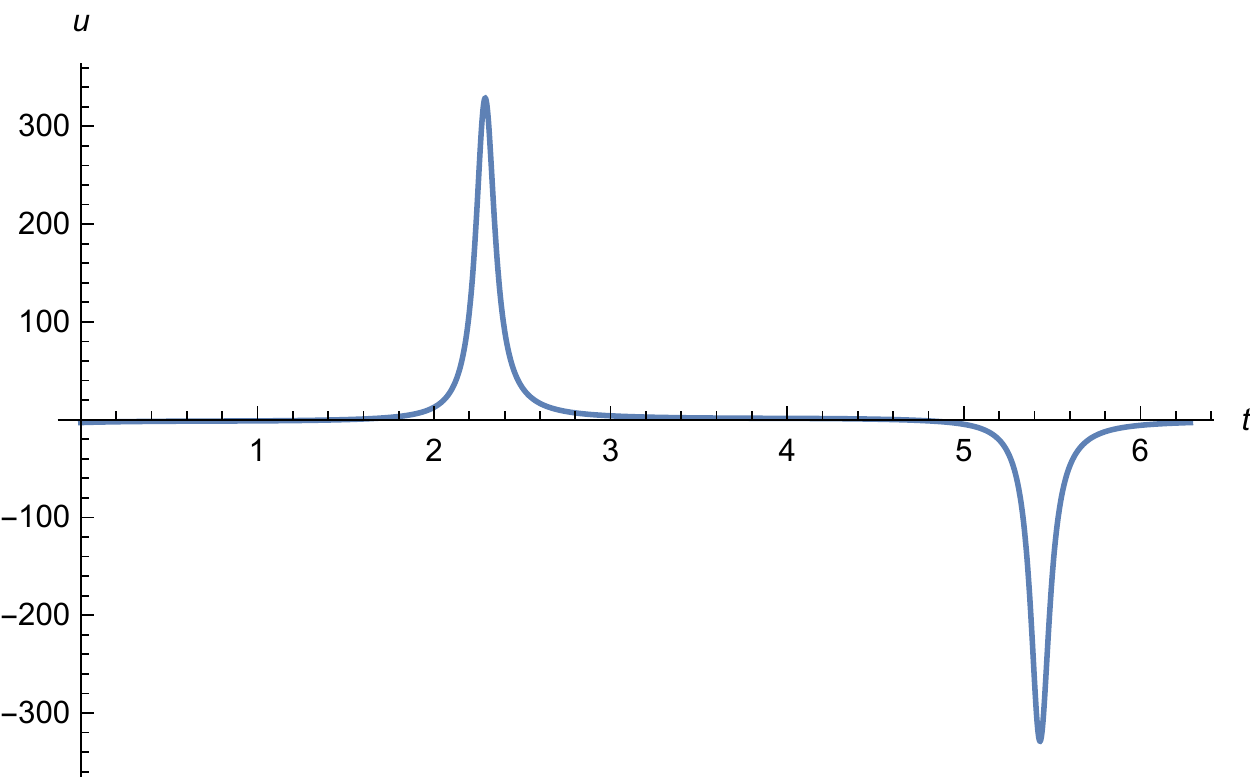}                  
     \end{center}
    \caption{$u(t, x=2.26)$ for the defocusing MKdV.}        
 \label{defocusing}
    \end{figure}

   The third argument in favor of the conjecture that the $x$ evolution of sufficiently smooth  Cauchy data  on the line $x = const$ for the MKdV equation \p{MKdV} does not bring about singularities in $u(t,x)$ comes from numerical simulations.
 To simplify the numerical analysis, we considered the problem on the band $0 \leq t \leq 2\pi$, where we imposed {[}as is allowed by Eq. \p{MKdV}{]}
periodic boundary conditions: 
 \be
        \lb{period}
        u(t + 2\pi, x) \ =\ u(t,x) \,.
\ee
 We have chosen 
the Cauchy data
\be \lb{cauchynum}
  u(t,0) \ =\ \sin t, \ u_x(t, 0) \ =\ u_{xx}(t,0) \ =\ 0\,.
\ee
We first checked that the use of such Cauchy data for the defocusing MKdV equation \p{MKdVm} was leading  to a blow-up rather fast  (at $x=2.2630\ldots$).
This is illustrated in Fig. \ref{defocusing}, where the function $u(t,x)$ is plotted just before the blow-up, at $x = 2.26$.

 By contrast, our numerical simulations of the $x$ evolution
of the focusing MKdV equation showed that $u(t,x)$ stayed bounded for all the values of $x$ that we explored. We met, however, another problem associated with the {\it instability} of Eq. \p{MKdV} under high-frequency (HF) perturbations.

Suppressing the nonlinear term in the  KdV or MKdV equations, we obtain
\be \lb{lKdV} 
 u_{xxx} + u_t   = 0 \,,
 \ee
This equation describes the fluctuations around the solution $u(t,x) = 0$. Its analysis gives us an idea on the behaviour of fluctuations around other solutions.
Decomposing $u(t,x)$ as a Fourier integral, 
 in  plane waves $ e^{ {\rm i} (\omega t + k x)}$, we obtain  
 the dispersion law
 \be \lb{disp}
 \omega = k^3\,.
 \ee
 If one poses the conventional Cauchy problem 
with some Fourier-transformable initial data
 \be
 u(0,x) = v(x) \equiv \int \frac{d k}{2 \pi} \, v(k) e^{ {\rm i}  k x}\,,
 \ee
 the time evolution of the initial data $v(x)$ yields the solution
 \be
 u(t,x) = \int \frac{d k}{2 \pi} \, v(k) e^{ {\rm i}(k^3t+  k x)}\,.
 \ee 
 The important point here is that $u(t,x)$ is obtained from $v(k)$ by a purely oscillatory complex
 kernel  $e^{ {\rm i}(k^3t +  k x)}$ of unit modulus. It has been shown that this oscillatory
 kernel has {\it smoothing} properties (see, e.g., \cite{Kenig1991}). This allows one to take
 the initial data in low-$s$ Sobolev spaces $H^s$ (describing pretty rough initial data).
 
{On the other hand}, if one considers the $x$-evolution Cauchy problem, one starts from three independent
functions of $t$ along the $x=0$ axis: $u(t,0)= u_0(t)$,  $u_x(t,0)= u_1(t)$ and $u_{xx}(t,0)= u_2(t)$, as in 
 \p{cauchynum}.
Assuming that the three Cauchy data $u_a(t)$, $a=0,1,2$, are Fourier-transformable, we can represent them as
 \be
 u_a(t)  \ =\ \int \frac{d \omega}{2 \pi} \, u_a(\omega) e^{ {\rm i}  \omega t}\,.
 \ee
 The three Cauchy data determine a unique solution which, when decomposed
 in plane waves, satisfies the same dispersion law \p{disp} as before. However,
 the dispersion law \p{disp} must now be solved for $k$ in terms of  $\omega$.
 As it is a cubic equation in $k${,} it has {\it three different roots}: 
 \be
 k_a(\omega) \ = \ \omega^{\frac13} e^{2\pi i a/3} \, \qquad a = 0,1,2\,.
  \ee
 This yields a solution for $u(t,x)$ of the form
\be
u(t,x)=\sum_{a=0,1,2} \int \frac{d \omega}{2 \pi} \, v_a(\omega) e^{ {\rm i} ( \omega t+ k_a x)}\,,
\ee
where the three coefficients $v_a(\omega)$ are uniquely determined by the three initial conditions
at $x=0$. 
The point of this exercise was to exhibit the fact that, when considering the $x$ evolution with arbitrary
Cauchy data $u_0(t)$,  $ u_1(t)$, $u_2(t)$, the solution involves {\it exponentially growing} modes
in the $x$ direction, linked to the imaginary parts of $k_1(\omega)$ and $k_2(\omega)$.

  \begin{figure} [ht!]
      \begin{center}
    \includegraphics[width=.5\textwidth]{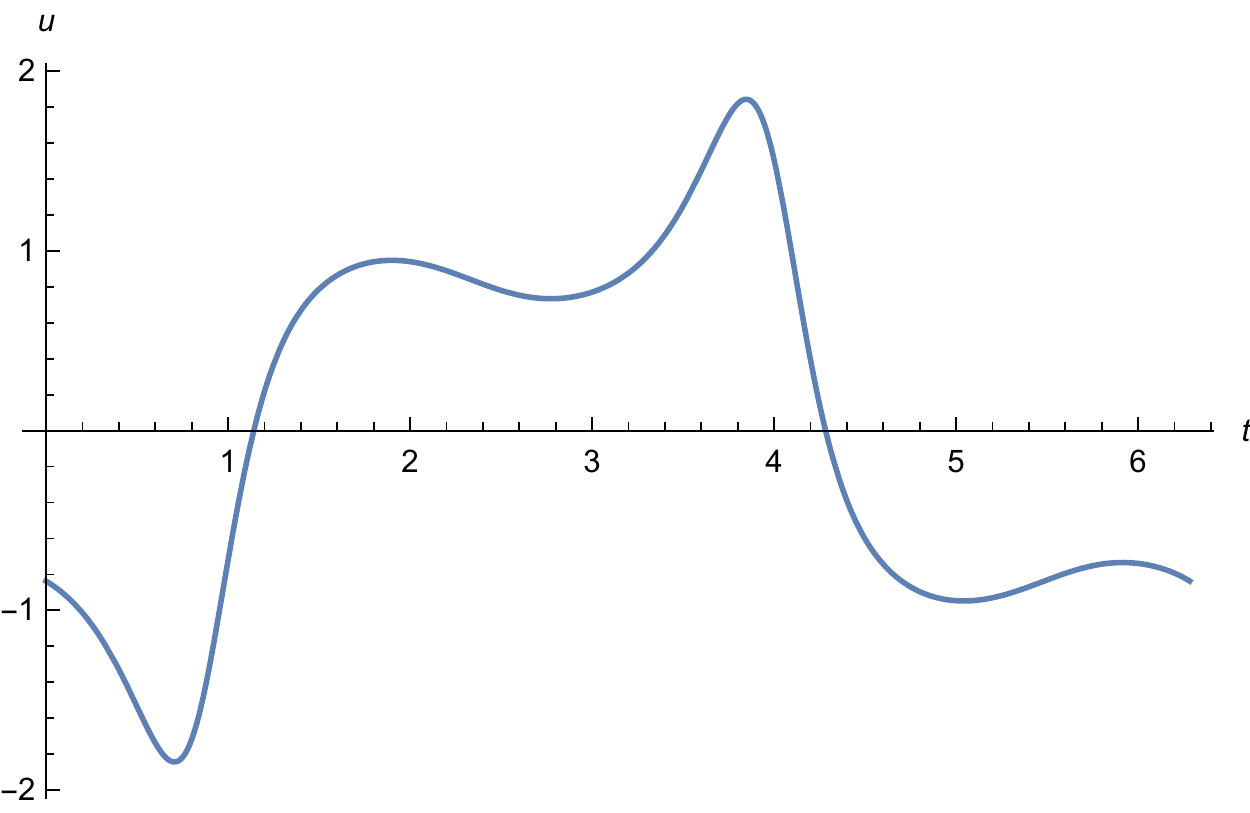}                  
     \end{center}
    \caption{$u(t, x=3)$ for the focusing MKdV.}        
 \label{focusing-u}
    \end{figure}    

\begin{figure} [ht!]
      \begin{center}
    \includegraphics[width=.5\textwidth]{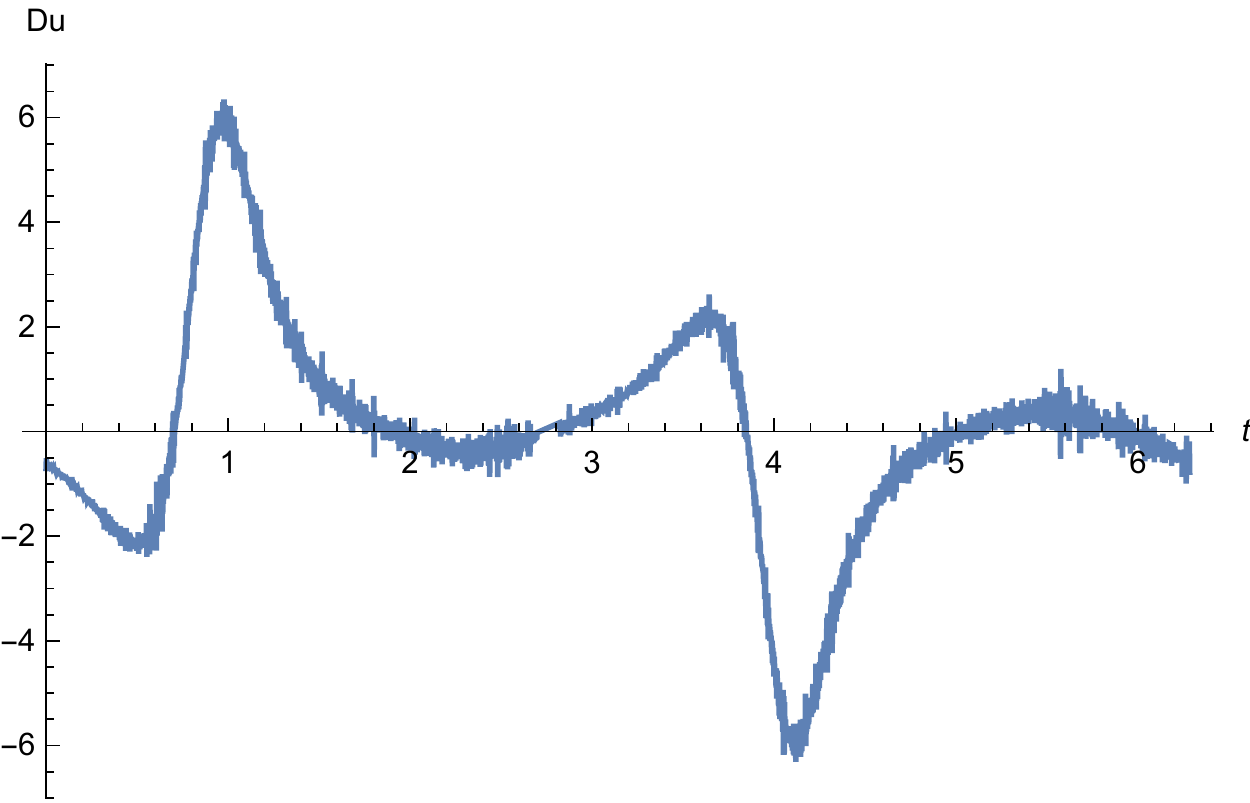}                  
     \end{center}
    \caption{$u_x(t, x=3)$ for the focusing MKdV.}        
 \label{focusing-Du}
    \end{figure}    

This can be avoided if the initial data are sufficiently smooth, not involving   HF  modes. 
As a minimum condition for a local existence theorem, one 
 should require the Fourier transforms $v_a(\omega)$ to decrease like 
$e^{- \alpha |\omega|^{\frac13} }$ for some positive constant $\alpha$.\footnote{See Ref.\cite{Damour} for more detailed discussion.}  

However, it is difficult to respect these essential smoothness constraints on the behavour of $u(t,x)$ in the numerical calculations. The standard Mathematica algorithms do not do so, and that is why we, starting from some values of $x$, observe the HF noise in our results. 

 In Figs. \ref{focusing-u},\ref{focusing-Du}, we present the results of numerical calculations of $u(t,x)$ and $u_x(t,x)$ for $x=3$. There is no trace of blow-up. For the plot of $u(t,x)$, one also does not see  a HF noise, but it is seen in the plot for $u_x(t,x)$.  For larger values of $x$, the noise also shows up in the plot of  $u(t,x)$.  At $x \ga 3.8$, the noise overwhelms the signal.

The observed noise is a numerical effect associated with a finite computer accuracy. To confirm this, we performed a different calculation choosing the initial conditions which correspond to the exact solitonic solution to Eq. \p{MKdV}.

The soliton is a travelling wave, $u(t,x) = u(x - ct) \equiv u(\bar x)$. Plugging this Ansatz into \p{MKdV}, we obtain an ordinary differential equation
\be
\frac{\partial}{\partial \bar x} \left[u_{\bar x \bar x} + 2 u^3 - cu  \right]=0\,.
\ee
Denoting as $C'$ the constant quantity within the bracket, we then get the following second-order equation
for the function $u(\bar x)$:
\be
 u_{\bar x \bar x}= - \frac{d}{d u} {\cal V}(u)\,,
\ee
with a potential function ${\cal V}(u)$ now given by
\be
{\cal V}(u)\ = \ \frac {u^4 -  c u^2}2 - C' u\,.
\ee
As was also the case for the time-independent Ansatz, the problem is reduced to the dynamics of a particle moving in the confining
quartic potential
${\cal V}(u)$. 
The trajectory of the particle depends 
on three parameters: the celerity $c$, the constant $C'$, and the particle energy,
\be
E = \frac12 u_{\bar x}^2 + {\cal V}(u)\,.
\ee
The usually considered solitonic solutions (such that $u(\bar x)$ tends to zero when $\bar x \to \pm \infty$)
are obtained by taking $c >0$, $C'=0$ (so that the potential represents a symmetric double-well potential) and $E=0$.  The zero-energy trajectory describes a particle starting at ``time" $\bar x = - \infty$, 
at $u=0$ with zero ``velocity" $u_{\bar x}$,
 gliding down, say, to the right, reflecting on the right wall of the double well and then turning back
to end up again at  $u=0$ when  $\bar x = + \infty$. The explicit form of the corresponding solution defined on the infinite $(t,x)$ plane is 
\be 
u(t,x) \ =\   \frac {\sqrt{c}} {\cosh [\sqrt{c} \, (x - ct)] } \,.
 \ee

\begin{figure} [ht!]
      \begin{center}
    \includegraphics[width=.5\textwidth]{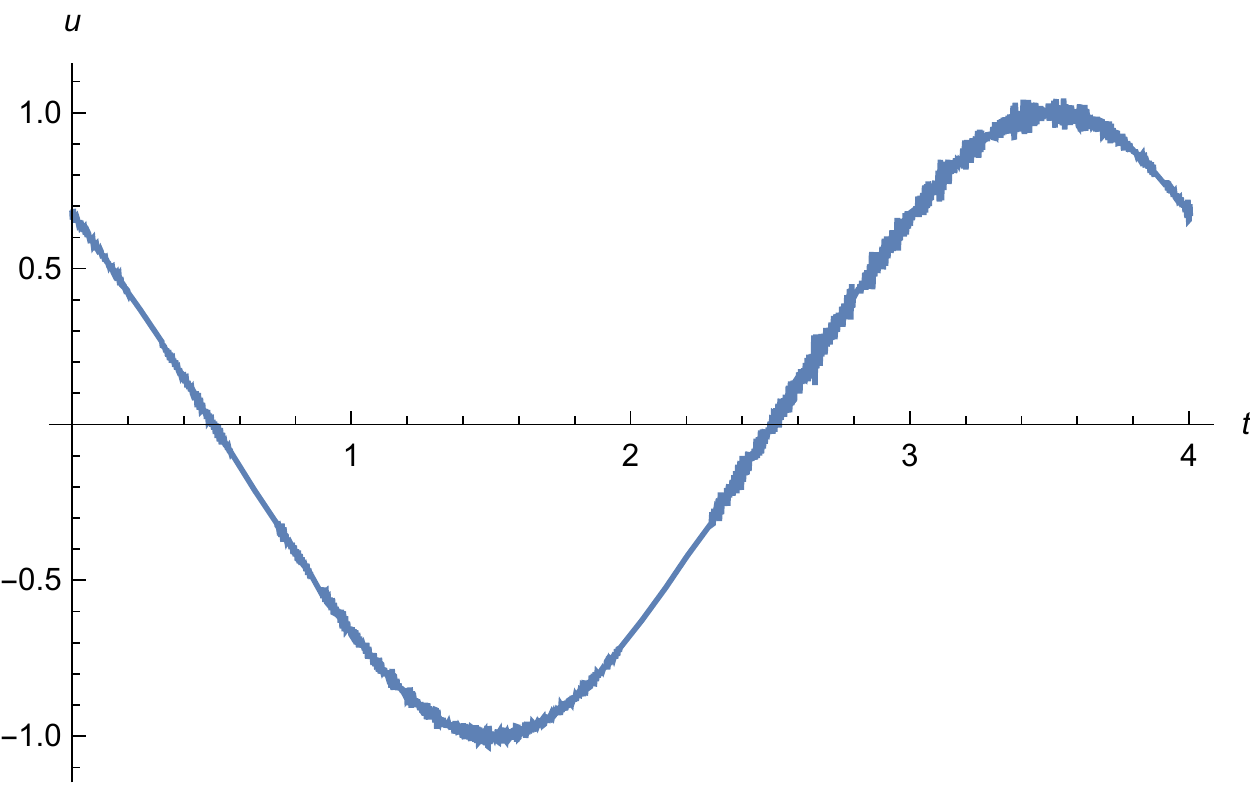}                  
     \end{center}
    \caption{HF noise for the periodic soliton evolution. $x = 4.5$.}        
 \label{persol}
    \end{figure}

  To make contact with our numerical calculations, we need, however, a {\it periodic} soliton solution.  Such solutions can be easily constructed by considering bound mechanical motions in the potential ${\cal V}(u)$  having a 
{\it non-zero energy}. Periodic solutions exist both for  positive and negative $c$. The trajectories are the elliptic functions. It was more convenient for us to assume $c = -|c|$, in which case we could make contact with  Ref. \cite{Robert}, where the expressions for the trajectories of motion in the same quartic potential were explicitly written, one only had to rename the parameters. Choosing $E=1$ and $c = -1$, we obtain the following solution:
 \be
\lb{cn}
 u(t,x) \ =\  {\rm cn} \left[\sqrt{3} (x+t), m \right]\,,
 \ee
where cn$(z)$ is the Jacobi elliptic cosine function with the elliptic modulus $m = 1/3$. The function \p{cn} is periodic both in $t$ and $x$ with the period
 \be T \ =\ L \ = \frac 4 {\sqrt{3}} K\left( \frac 13 \right) \ \approx 4\,.\ee
We fixed the initial conditions for $x=0$ and periodic conditions in time as is dictated by \p{cn} and then numerically solved \p{MKdV}. The numerical solution should reproduce the exact one, and it does for $x \la 4$. However,  at larger values of $x$, the HF noise appears. The result of the calculation for $x=4.5$ is given on Fig. \ref{persol}.

One can suppress the HF noise by increasing the step size, but then the form of the soliton is distorted. To find a numerical procedure that suppresses the noise and gives correct results for large values of $x$ remains a challenge for future studies.

   \section{Discrete models with benign ghosts}

One of the possible solutions to this numerical problem could consist in discretizing the model in time direction and 
assuming that the variable $t$ takes only the discrete values $t=h, 2 h, \cdots, N h$, 
for some integer $N \geq 3$
 and by replacing the continuous time derivative $\psi_t$ by a discrete (symmetric) time derivative 
 $[\psi(t+h,x)-\psi(t-h,x)]/(2 h)$.
Then the Lagrangian\footnote{It is quite analogous to \p{LKdV}.  After variation with respect to $\psi(t,x)$, one gets Eq.  \p{MKdV} after posing $ u(t,x) = \psi_x(t,x)$.}
 \be
      \lb{LMKdV}
      L[\psi(t,x)] \ =\ \frac {\psi_{xx}^2 - \psi_x^4 -   \psi_x \psi_t}2\,.
       \ee
acquires the form 
   \be \label{LN}
L_N= \sum_{k=1}^{N} \left\{ \frac { [\psi_{xx}(kh,x)]^2 -  [\psi_x(kh,x)]^4}2 - \right. \nn
\left.
\frac 12  \psi_x(kh,x) \frac{\psi[(k+1)h,x]-\psi[(k-1)h,x] }{2 h}  \right\}\,,
 \ee   
where we impose the periodicity: $\psi(0,x) \equiv  \psi(Nh, x)$ and  $\psi[(N+1)h,x] \equiv  \psi(h, x)$.\footnote{It is also possible to  impose the Dirichlet-type boundary conditions,    $\psi(0h,x)= 
 \psi[(N+1)h,x]=0$. For $N=2$, periodicity cannot be imposed and Dirichlet conditions is the only option.} 

The Lagrangian \p{LN} includes a finite number of degrees of freedom and represents a mechanical system. This system involves higher derivatives in $x$  (playing the role of time) and hence involves ghosts. 
Defining the  new dynamical variables      $a^k(x)  =  \psi_x(kh, x)$,   the equations
of motion derived from the Lagrangian     \p{LN} read
      \be 
\lb{discr-eq}
       a^k_{xxx} + 6 (a^k)^2 a^k_x  \ + \  \frac {a^{k+1} - a^{k-1}}{2h}\ =\ 0\,.
        \ee
There are two integrals of motion: the energy
  \be \label{EN}
        E \ =\ \sum_{k=1}^N \left[ \frac { (a^k_x)^2 - 3(a^k)^4}2  - a^k a^k_{xx}  \right]\,.
         \ee  
and 
\be \label{QN}
 Q \ =\ \sum_{k=1}^N \left[ a^k_{xx} + 2 (a^k)^3  \right]\,.
  \ee
 The expression \p{EN} is the discretized version of the 
integral \be
\int dt \left[\frac{(u_x)^2 - 3u^4}2  - u u_{xx}\right]
\ee
 in the continuous MKdV system, which is conserved during the $x$ evolution, as follows from the local conservation law \p{Energy}. The second integral of motion is related to the conservation law \p{protoEnergy}. By contrast, the currents  in the higher conservation laws of  the MKdV equation,
 starting with Eq.\p{superEnergy}, do not translate into integrals of motion of the discrete systems. We have only two integrals of motion and many variables, which means that the equation system \p{discr-eq} is not integrable and exhibits a chaotic behaviour.

We fed these equations to Mathematica and found out that their solution stays bounded up to $x = 10000$ and more --- the ghosts are benign! This represents a further argument  in favour of the conjecture that also in the continuous theory the evolution in spatial direction is benign. Indeed, one may  
 expect that taking larger and larger values of $N$ would allow one to simulate better and better  the 
continuous theory (though the presence of chaos might make such a convergence non uniform in $x$).

\begin{figure} [ht!]
      \begin{center}
    \includegraphics[width=.5\textwidth]{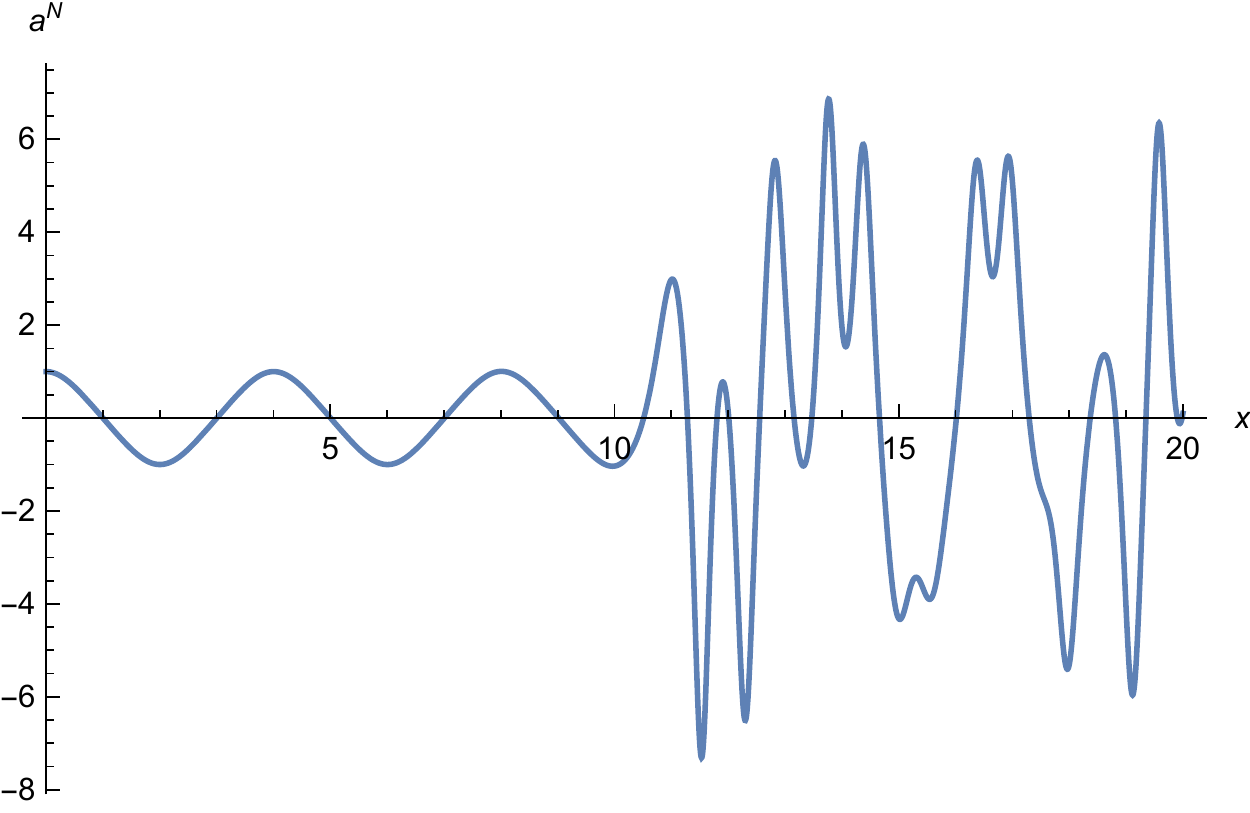}                  
     \end{center}
    \caption{The solution of the system \p{discr-eq} for $a^N(x)$ ($N = 350, \  h =T/N)$.}        
 \label{discrete}
    \end{figure}

Anyway, we tried the solitonic initial conditions and found out that the discrete system for large $N=350$ (the limit of Mathematica skills) behaves better than the PDE. As is seen from Fig. \ref{discrete}, the discrete solution stays close to the exact soliton solution up to $x \approx 10$, to be compared to $x \approx 4.5$, which was the horizon of the numerical procedure of the previous section. Hopefully, a clever mathematician, an expert in numerical calculations, would be able to increase the horizon still more...

Lastly, we note that, irrespectively to the relationship of the systems \p{LN} to the MKdV equation, these systems represent an interest by their own because they provide a set of nontrivial interacting higher derivative systems with benign ghosts. Such systems were not known before.

          \section*{Acknowledgements}

         I am grateful to the organizers of the AAMP conference for the invitation to make a talk there. Working on our paper \cite{Damour}, we benefited a lot from  illuminating discussions with Piotr Chrusciel, Alberto De  Sole,  Victor Kac, Nader Masmoudi, Frank Merle and Laure Saint-Raymond.

\end{document}